\documentclass{elsart}

\usepackage{cite}

\begin{document}

\begin{frontmatter}

\title{About homotopy perturbation method for solving heat--like
and wave--like equations with variable coefficients}

\author{Francisco M. Fern\'{a}ndez \thanksref{FMF}}

\address{INIFTA (UNLP, CCT La Plata-CONICET), Divisi\'{o}n Qu\'{i}mica Te\'{o}rica,\\
Diag. 113 y 64 (S/N), Sucursal 4, Casilla de Correo 16,\\
1900 La Plata, Argentina}

\thanks[FMF]{e--mail: fernande@quimica.unlp.edu.ar}

\begin{abstract}
We analyze a recent application of homotopy perturbation method to
some heat--like and wave--like models and show that its main results are
merely the Taylor expansions of exponential and hyperbolic functions.
Besides, the authors require more boundary conditions than those already
necessary for the solution of the problem by means of power series.
\end{abstract}

\end{frontmatter}

\section{Introduction}

In the last years there has been great interest in the power--series
solution of physical problems by means of homotopy perturbation method\cite
{CHA07,RDGP07,YO07,SG08,OA08}. In a series of papers I have shown that the
homotopy perturbation method, the Adomian decomposition method and the
variational iteration method have produced the poorest and most laughable
papers ever published\cite{F07,F08b,F08,F08c,F08d,F08e,F08f}. Notice, for
example the application of homotopy perturbation method to obtain the Taylor
expansion of exponential functions of the form $e^{iat}$\cite{SG08}.
Curiously, some journals believe that such contributions may be of interest
for the scientific community.

The purpose of this article is to discuss a recent paper by \"{O}zi\c{s} and
A\u{g}\i rseven\cite{OA08} who have recently applied the homotopy
perturbation method to heat--like and wave--like equations. We compare their
approach with the straightforward power--series one.

\section{Heat--like models}

\"{O}zi\c{s} and A\u{g}\i rseven studied particular cases of the equation%
\cite{OA08}
\begin{equation}
\frac{\partial u(\mathbf{r},t)}{\partial t}=\hat{L}u(\mathbf{r},t)+f(\mathbf{%
r})  \label{eq:heat}
\end{equation}
where $\hat{L}$ is a linear operator and $f(\mathbf{r)}$ a differentiable
function. As a first approximation we can naively try a time--power--series
solution of the form
\begin{equation}
u(\mathbf{r,}t)\mathbf{=}\sum_{j=0}^{\infty }u_{j}(\mathbf{r})t^{j}
\label{eq:u_series}
\end{equation}
If we substitute equation (\ref{eq:u_series}) into equation (\ref{eq:heat})
we obtain a recurrence relation for the coefficients of the expansion (\ref
{eq:u_series}):
\begin{equation}
u_{n+1}=\frac{1}{n+1}\left( \hat{L}u_{n}+f\delta _{n0}\right)
,\,n=0,1,\ldots   \label{eq:u_n_1}
\end{equation}
where $\delta _{ij}$ is the Kronecker delta. Our first observation is that
we only need the coefficient
\begin{equation}
u_{0}(\mathbf{r})=u(\mathbf{r},0)  \label{eq:u0}
\end{equation}
in order to obtain the remaining ones by means of the recurrence relation (%
\ref{eq:u_n_1}).

The first example studied by \"{O}zi\c{s} and A\u{g}\i rseven\cite{OA08} is
a particular case of equation (\ref{eq:heat}) with
\begin{equation}
\hat{L}=\frac{x^{2}}{2}\frac{\partial ^{2}}{\partial x^{2}},\,f=0
\label{eq:ex1}
\end{equation}
If we apply the recurrence relation (\ref{eq:u_n_1}) with $%
u_{0}(x)=u(x,0)=x^{2}$ we obtain the coefficients of the time--power series
for the exact solution
\begin{equation}
u(x,t)=x^{2}e^{t}  \label{eq:u_ex1}
\end{equation}
that one easily obtains by the method of separation of variables. It is
amazing that \"{O}zi\c{s} and A\u{g}\i rseven\cite{OA08} require the
additional boundary conditions $u(0,t)=0$ and $u(1,t)=e^{t}$. On the other
hand, the power--series method reveals that the initial condition completely
determines the solution obtained by those authors and that the remaining
boundary conditions are redundant. But the most important fact is that those
authors were allowed to publish an elaborated perturbation method for
obtaining the Taylor series of $e^{t}$!!!\cite{OA08}.

The reader may easily verify that the straightforward application of the
recurrence relation (\ref{eq:u_n_1}) enables one to solve all the other
heat--like equations chosen by \"{O}zi\c{s} and A\u{g}\i rseven\cite{OA08},
and in all the cases one needs \textit{only} the initial condition (\ref
{eq:u0}) in order to obtain the solutions. On the other hand, \"{O}zi\c{s}
and A\u{g}\i rseven\cite{OA08} require some additional boundary conditions.
For completeness we outline the results in what follows:

Example 2:
\begin{equation}
\hat{L}=\frac{1}{2}\left( y^{2}\frac{\partial ^{2}}{\partial x^{2}}+x^{2}%
\frac{\partial ^{2}}{\partial y^{2}}\right) ,\,f=0  \label{eq:ex2}
\end{equation}
The authors require four Neumann boundary conditions and the initial one $%
u(x,y,0)=y^{2}$\cite{OA08}. The power series method completely determines
the exact solution $u(x,y,t)=y^{2}\cosh t+x^{2}\sinh t$ from only the
initial condition. Notice that in this case the authors were able to obtain
the Taylor series for $\sinh t$ and $\cosh t$ which are far more difficult
than the preceding feat.

Example 3:
\begin{equation}
\hat{L}=\frac{1}{36}\left( x^{2}\frac{\partial ^{2}}{\partial x^{2}}+y^{2}%
\frac{\partial ^{2}}{\partial y^{2}}+z^{2}\frac{\partial ^{2}}{\partial z^{2}%
}\right) ,\,f=x^{4}y^{4}z^{4}  \label{eq:ex3}
\end{equation}
The authors require six Neumann boundary conditions and the initial one $%
u(x,y,z,0)=0$\cite{OA08}. The time--power series completely determined by
the initial condition clearly converges towards the exact solution $%
u(x,y,z,t)=x^{4}y^{4}z^{4}(e^{t}-1)$. In this case the authors clearly show
that the homotopy perturbation method is able to provide the Taylor series
for $e^{t}-1$ about $t=0$.

\section{Wave--like models}

\"{O}zi\c{s} and A\u{g}\i rseven\cite{OA08} also studied some wave--like
equations that are particular cases of
\begin{equation}
\frac{\partial ^{2}u(\mathbf{r},t)}{\partial t^{2}}=\hat{L}u(\mathbf{r},t)+f(%
\mathbf{r})  \label{eq:wave}
\end{equation}
If we substitute the time--power series (\ref{eq:u_series}) we realize that
one can obtain the coefficients from the recurrence relation
\begin{equation}
u_{n+2}=\frac{1}{\left( n+1\right) \left( n+2\right) }\left( \hat{L}%
u_{n}+f\delta _{n0}\right) ,\,n=0,1,\ldots   \label{eq:u_n_2}
\end{equation}
In this case if we have the first two coefficients
\begin{equation}
u_{0}(\mathbf{r})=u(\mathbf{r},0),\,u_{1}(\mathbf{r})=\left. \frac{\partial
u(\mathbf{r},t)}{\partial t}\right| _{t=0}  \label{eq:u_0_u_1}
\end{equation}
we can obtain the remaining ones. The example 4 studied by \"{O}zi\c{s} and A%
\u{g}irseven\cite{OA08} is a particular case of equation (\ref{eq:wave})
with the linear operator and function given by equation (\ref{eq:ex1}). If
we choose the coefficients $u_{0}(x)=x$ and $u_{1}(x)=x^{2}$ proposed by
\"{O}zi\c{s} and A\u{g}\i rseven\cite{OA08} we easily obtain the remaining
coefficients of the time--power series of the exact solution
\begin{equation}
u(x,t)=x+x^{2}\sinh (t)  \label{eq:u_ex4}
\end{equation}
Once again we see that we do not need the other boundary conditions $u(0,t)=0
$, and $u(1,t)=1+\sinh (t)$ already required by \"{O}zi\c{s} and A\u{g}%
irseven\cite{OA08}. In this case the authors were able to produce the Taylor
series for $\sinh (t)$.

We draw the same conclusion regarding the other wave--like models studied by
\"{O}zi\c{s} and A\u{g}\i rseven\cite{OA08} as shown in what follows:

Example 5:
\begin{equation}
\hat{L}=\frac{1}{12}\left( x^{2}\frac{\partial ^{2}}{\partial x^{2}}+y^{2}%
\frac{\partial ^{2}}{\partial y^{2}}\right) ,\,f=0  \label{eq:ex5}
\end{equation}
The authors require four Neumann conditions in addition to the initial ones $%
u_{0}(x,y)=x^{4}$, and $u_{1}(x,y)=y^{4}$\cite{OA08}. One easily verifies
that the straightforward power--series method yields the exact solution $%
u(x,y,t)=x^{4}\cosh t+y^{4}\sinh t$ from only the two initial conditions
already indicated. Once again, the authors were able to obtain the Taylor
series for $\cosh t$ and $\sinh t$.

Example 6:
\begin{equation}
\hat{L}=\frac{1}{2}\left( x^{2}\frac{\partial ^{2}}{\partial x^{2}}+y^{2}%
\frac{\partial ^{2}}{\partial y^{2}}+z^{2}\frac{\partial ^{2}}{\partial z^{2}%
}\right) ,\,f=x^{2}+y^{2}+z^{2}  \label{eq:ex6}
\end{equation}
The authors made use of six boundary conditions and the two initial ones $%
u_{0}(x,y,z)=0$, and $u_{1}(x,y,z)=x^{2}+y^{2}-z^{2}$\cite{OA08}. In this
case we verify that the power--series method with the two initial conditions
completely determine the exact solution $%
u(x,y,z,t)=(x^{2}+y^{2})(e^{t}-1)+z^{2}(e^{-t}-1)$. As the reader may have
already guessed, \"{O}zi\c{s} and A\u{g}\i rseven\cite{OA08} were able to
derive the Taylor expansions for $e^{t}-1$ and $e^{-t}-1$ about $t=0$.

\section{Conclusions}

We have seen that the homotopy perturbation method proposed by \"{O}zi\c{s}
and A\u{g}\i rseven\cite{OA08} produce the Taylor series about $t=0$ of the
solutions of some differential equations. Curiously, those authors require
more boundary conditions than the ones already necessary for the
straightforward application of the power--series approach. The main results
in their paper are Taylor expansions of exponential and hyperbolic functions
of time about $t=0$. It is amazing that a supposedly respectable journal
accepts such contribution. However, I may be mistaken and in a near future
all the courses on calculus will be teaching homotopy perturbation theory
instead of Taylor series.

\end{document}